\documentstyle[12pt]{article}
\begin{document}
\title{Hard Diffraction in the QCD Dipole
Picture}
\author{A.Bialas and R.Peschanski\\Institute of Physics, Jagellonian
University\\ Reymonta 4,  30-059 Cracow, Poland\\CEA, Service de Physique
Theorique, CE-Saclay\\ F-91191 Gif-sur-Yvette Cedex, France}
\maketitle
\begin{abstract}
Using the QCD dipole picture of the BFKL pomeron, the gluon
contribution to the cross-section
for single
 diffractive dissociation in deep-inelastic high-energy  scattering is calculated.
The resulting contribution to the proton diffractive structure function
integrated over $t$ is given in
terms of relevant variables, $x_{\cal P}, Q^2, $ and $\beta = x_{Bj}/x_{\cal P}.$
It  factorizes into an explicit $x_{\cal P}-$dependent
Hard Pomeron flux factor and
structure function.
The flux factor is found to have substantial logarithmic corrections
which may account for the recent measurements of the Pomeron intercept in
this process. The triple Pomeron coupling is shown to be strongly enhanced 
by the resummation of leading logs.
The obtained pattern of scaling violation at small $\beta$ is
similar to that for $F_2$ at small $x_{Bj}.$
\end{abstract}

{\bf 1.} Recently, new measurements of the proton diffractive structure function at
small $x$ and very large $Q^2$ were presented by H1 and ZEUS experiments
\cite{h1,zeus}.  The observed   3-dimensional
structure function factorizes:
\begin{equation}
F_2^{D(3)}(x_{Bj},Q^2,x_{\cal P}) = f(x_{\cal P}) \ F_2^{D(2)}(\beta, Q^2)\ .
\label{F3}
\end{equation}
Here $\beta \equiv  Q^2/
(Q^2 + M^2),$ $x_{\cal P} = x_{Bj}/\beta,$ and
$M^2$ is the mass of the diffractively excited system.
This factorized form is naturally interpreted as the product of a Pomeron
flux factor inside the proton and its corresponding structure function\cite{Ingelman}. However, other theoretical interpretations
are also possible\cite{theory}.

The aim of the  present paper is to investigate the perturbative QCD
contribution to the process in question using the colour
 dipole approach \cite{muel,nik}, which is known \cite{muel,nik,muela} to
reproduce the physics
of the "Hard Pomeron"\cite{lip}. We calculate the diffraction dissociation of
the virtual photon on the proton  at small $x_{Bj}$ and large $Q^2.$
Our main result is  the explicit formula for the small $x_{\cal P}$ behaviour
of the 3-dimensional diffractive structure function
\begin{displaymath}
 F_2^{D(3)}(Q^2,x_{\cal P},\beta) \equiv \int dt F_2^{D(4)}(Q^2,x_{\cal
P},\beta,t) =
\end{displaymath}
\begin{equation}
 =
 \frac{2 e^2_f \alpha^5 N_c^2}{\pi^2} \left(\frac{2a(x_{\cal P})}{\pi}\right)^3
x_{\cal P}^{-1-2\Delta_{\cal P}}
 \int_{c-i\infty}^{c+i\infty}\frac{d\gamma}{2\pi i}\
\left(\frac{r_0Q}2 \right)^{\gamma}
 H(\gamma)\ 
\beta^{-\alpha N_c \chi(\gamma)/\pi} \label{Fdfin}
\end{equation}
where $H(\gamma)$ is given\footnote {We consider here only the transverse photon contribution.} by
\begin{equation}
H(\gamma)=V(\gamma) \frac 4{\gamma ^2 (2-\gamma)^4} \frac{ \Gamma(3-\frac{\gamma}2)
\Gamma^3(2-\frac{\gamma}2)
 \Gamma(2+\frac{\gamma}2)
\Gamma(1+\frac{\gamma}2)}{\Gamma(4-\gamma)
\Gamma
(2+\gamma)
}
\label{hnik}
\end{equation}
with
\begin{equation}
V(\gamma)=
\ \int_{0}^{1} F(1-\frac{\gamma}2,
1-\frac{\gamma}2;1;y^2)\ dy
\label{V0}
\end{equation}
(F is the hypergeometric function).
 $\chi(\gamma)$ is the eigenvalue of the BFKL kernel defined as
\begin{equation}
\chi(\gamma) = 2\psi(1) - \psi(1-\frac{\gamma}{2})-\psi(\frac{\gamma}{2}),
\label{chi}
\end{equation}
and  $a(\xi)$
is given by
\begin{equation}
    a(\xi) = [7\alpha N_c\zeta(3)\log (1/\xi)/\pi]^{-1}.      \label{ax}
\end{equation}
$\Delta_{\cal P} \equiv \alpha_{\cal P} -1 =
\frac{\alpha N_c}{\pi} \chi (1)\ $.
 $r_0$ is a non-perturbative parameter (defined by (\ref{r0}) as the
average of the transverse dipole size inside the target) which
cannot be determined within the present approach. Finally,
$e_f^2$ is the sum of the squares of quark charges.

The formula (\ref{Fdfin}) has several interesting features.

(i) {\sl Factorization}. One sees that $F_2^{D(3)}$ is a product of two
factors: one depends only on $x_{\cal P}$,
another one depends on
$\beta$ and
$Q^2$. The first one
can thus  be interpreted as the
"Pomeron flux
factor" and the second one as the "Pomeron structure function" \cite{h1,zeus}.

(ii) {\sl Pomeron flux factor}. With this identificaton, one
obtains for the Pomeron flux factor
\begin{equation}
\Phi_{\cal P} =  C_{\Phi}\  x_{\cal P}^{-1-2\Delta_{\cal P}}
\left(\frac{2a(x_{\cal P})}
{\pi} \right)^3 \label{flux}
\end {equation}
where $C_{\Phi}$ is an arbitrary constant.
One sees that the obtained  $x_{\cal P} $ dependence differs from the normally
assumed
power law $x_{\cal P}^{-1-2\Delta_{\cal P}} $ by a logarithmic factor $\left(
\log (1/x_{\cal P}) \right)^{-3},$ as found already in the triple Pomeron
limit \cite{muela}. Our calculation shows that this logarithmic
correction is rather general and also applies beyond this limit. It would be of course very
interesting to verify this prediction with the data.
In this context we note that, when fitted with a power law, the formula
(\ref{flux}) gives an effective Pomeron intercept
\begin{equation}
 \Delta_{\cal P}^{eff} = \Delta_{\cal P} -\frac{3}{2
log(1/x_{\cal P})}. \label{eff}
\end{equation}
We would like to emphasize that this correction is  rather substantial even at
rather small
$x_{\cal P}$
(at
$x_{\cal P} = 10^{-3} , \Delta_{\cal P}^{eff} - \Delta_{\cal P}
\approx -0.2$). One sees that this effect agrees both in sign and in
magnitude with the  difference between the Pomeron intercepts observed in
$F_2$ \cite{zeusa,h1a} and in the diffractive structure function $F_2^{D(3)}$
\cite{h1,zeus} and
may thus be a simple explanation of the apparent contradiction between these
two measurements.
It would clearly be of great interest to analyze the future data using the form
(\ref{flux}).

(iii) {\sl The Pomeron structure function}
depends explicitly on
$Q^2$ and thus the
model predicts violation of scaling. To have a feeling on the pattern of this
scaling violation and also on the dependence of $F_{\cal P}$ on $\beta$, it is
illuminating to
evaluate the integral in (\ref{Fdfin}) by the saddle-point method. The result
is
\begin{equation}
 F_{\cal P}(Q^2,\beta)
 =
 \frac{ e^2_f \alpha^5 N_c^2}{C_{\Phi}\pi^2}
\left(\frac{r_0Q}2 \right)^{\gamma_0}
  H(\gamma_0)
\left(\frac{2a(\beta)7\zeta(3)}{\pi\chi''(\gamma_0)}\right)^{-\frac
12}
\beta^{-\alpha N_c \chi(\gamma_0)/\pi}, \label{fpsd}
\end{equation}
$\gamma_0$ being  determined from the saddle-point equation:
\begin{equation}
\frac{\alpha N_c}{\pi} \ \chi^{\prime} (\gamma_0) \ \log \beta = \log
(\frac{Qr_0}2).
\label{gamma0}
\end{equation}
      In the interesting "triple Pomeron" limit,
$\beta \approx 0, \gamma_0 \approx 1$, (\ref{fpsd}) simplifies into
\begin{equation}
 F_{\cal P}(Q^2,\beta\approx 0)
 =
 \frac{ e^2_f \alpha^5 N_c^2}{C_{\Phi}\pi^2} H(1)
\frac{r_0Q}2 \beta^{-\Delta_{\cal P}}
\left(\frac{2a(\beta)}{\pi}\right)^{\frac
12}  \exp[-\frac{a(\beta)}2 log^2(\frac{r_0Q}2)]. \label{3p}
\end{equation}
This formula gives a pattern of scaling    violation typical of the exchange of
a hard pomeron \cite{NPR}.

(iv) { \sl Triple Pomeron coupling}.
$H(\gamma)$ gives the triple pomeron coupling.
$H(1)$ can be explicitly evaluated from (\ref{hnik}) with the result $H(1)=
9 \pi^2 G /64$ where $G = .915...$ is  Catalan's constant. This value can be
compared with the corresponding coupling obtained from the same expressions
using the expansion $\gamma \! \rightarrow \! 0$, which corresponds to the
first-order perturbative QCD result \cite{catani}. One finds
$H^{pert}(\gamma) = \gamma^{-3}/12$. One may observe a large
enhancement
factor $H(1)/H^{pert}(1) = \frac {27 \pi ^2 G} 8 \approx 30$ due to the leading
$\log(1/x)$ resummation which is taken into account in the QCD dipole model. It provides  a theoretical
hint for the surprisingly
large experimentally observed hard-diffractive cross-section.

{\bf 2.} We shall now outline the derivation of the Eq.(\ref{Fdfin}). In order to
formulate the problem of diffraction dissociation we use the old idea of Good and Walker (see e.g.\cite{good,miet}), i.e. the expansion of the initial colliding state in  the diag
 onal basis of the eigenstates of absorption. To this end, we observe that the
dipole representation corresponds precisely to such a decomposition. Indeed, the amplitude for elastic scattering of two dipoles of transverse size
$x_1,x_2,$, is simply given by  two-gluon exchange\cite{muela,muelb}, namely:
\begin{equation}
T(x_1,x_2)= 4 \pi \alpha^2 \int \frac{dl}{l^3} [1-J_0(lx_1)] [1-J_0(lx_2)]\ ,
\label{T}
\end{equation}
and this interaction   changes neither their transverse size and position
nor  their rapidities.
Using this general
framework we write the  cross-section for
 single diffractive dissociation
\footnote {An analogous formulation was used in
\cite{muela}.}
 of a virtual photon on a proton as:
\begin{equation}
\frac{\beta d\sigma}{d\beta d^2b} =
\int d^2\bar{r} d\bar{z}\ \bar{\Phi} (\bar{r},\bar{z};Q^2) \\ \
\sigma_d(\bar{r},b,\beta,x_{\cal P})
\label{sigd}
\end{equation}
where  $\bar{\Phi}$ is the probability of the virtual photon
to fluctuate into a $q\bar{q}$ pair
 and $\sigma_d$ is the single diffractive
cross-section
in dipole-proton scattering
\eject
\begin{eqnarray}
\nonumber \sigma_d = \int  \frac{dx}{x} d^2\bar{s}_1 \
\hat{\rho}_1(b+\bar{s}_1,x,\xi)
\int
\frac{dx'}{x'} d^2\bar{s}_2\ 
\hat{\rho}_1(b+\bar{s}_2,x',\xi) \\  \int
\frac{d\bar{x}_1}{\bar{x}_1}
  \frac{d\bar{x}_2}{\bar{x}_2}
\ \rho_2(\bar{r};
\bar{s}_1,
\bar{x}_1; \bar{s}_2,\bar{x}_2;x_{Bj}/\xi,x_{\cal P}/\xi )
\ T(x,\bar{x}_1)\ T(x', \bar{x}_2)\ .
   \label{H}
\end{eqnarray}
$\rho_2$ is the  double dipole density
in the colliding dipole of transverse size $\bar{r}$ \cite{muela,muelb} and
$\hat{\rho}_1$ is the single dipole density in the proton:
\begin{equation}
\hat{\rho}_1(b,x,\xi) = \int d^2r dz \Phi(r,z) \rho_1(r,b,x,\xi) \label{hat}
\end{equation}
where $\Phi$ is the square of the proton wave function and  $\rho_1$ is the
single dipole density in a dipole of transverse size
$r$
\cite{muelb}
\begin{equation}
\rho_1(r,b,x,\xi) = \frac r{4xb^2} (2a(\xi)/\pi)^{\frac 32}
\log \left(b^2/rx\right)
 \xi^{-\Delta_{\cal P}} e^{ - \frac{a(\xi)}2 \log^2\left(b^2/rx\right)}\ .
\label{ro1}
\end{equation}

We start by computing the Mellin transform
\begin{equation}
 \tilde{\sigma}_d(\gamma;b;\beta,x_{\cal P}) \equiv \int d\bar{r}\
\bar{r}^{\gamma
-1} \sigma_d(\bar{r};b;\beta,x_{\cal P})/\bar{r}^2.
\label{mellin}
\end{equation}
Using the
 methods developed in
\cite{BP1}, one arrives at
the following
formula for the Mellin transform of the double-dipole density $\rho_2$:
\begin{eqnarray}
\nonumber \tilde{\rho}_2(\gamma;s_1,x_1,s_2,x_2;\xi_1,\xi_2) = \frac{2\alpha
N_c}{\pi} \
\left(\frac {\xi_2}{\xi_1}\right)^{\frac {\alpha N_c}{\pi} \chi(\gamma)} \\
\int \frac{dr_1}{r_1} \frac{dr_2}{r_2} \rho_1(r_1,b_1,x_1,\xi_2)\
\rho_1(r_2,b_2,x_2,\xi_2)\ W(r_1,r_2)
\label{ro2}
\end{eqnarray}
where, for $r_1<r_2$,  the symmetric function $ W(r_1,r_2)$  is given by
\cite{BP1}:
\begin{equation}
 W(r_1,r_2) = r_2^{\gamma -2}\ F(1-\gamma/2,1-\gamma/2;1;(r_1/r_2)^2)\
\label{W}
\end{equation}

By repeated application of the formula \cite{muelb}
\begin{eqnarray}
\nonumber \int \frac{dx_1}{x_1} \frac{dx_2}{x_2} T(x_1,x_2)
 d^2s\ \rho_1 (r_1,s,x_1,\xi_1) \rho_1 (r_2,b+s,x_2,\xi_2) = \\ \pi \alpha^2
 \frac{r_1 r_2}{b^2}(\xi_1
\xi_2)^{-\Delta_{\cal P}}   (\frac{2a(\xi_1\xi_2)}{\pi})^{3/2}
\log(\frac{b^2}{r_1
r_2}) \exp[-\frac{a(\xi_1 \xi_2)}{2} \log^2(\frac{b^2}{r_1 r_2})] \label{form}
\end{eqnarray}
one obtains
\begin{equation}
  \tilde{\sigma}_d = \frac{  \pi \alpha^5 N_c}{b^4 x_{\cal
P}^{\Delta_{\cal P}} }
\beta^{-\alpha N_c \chi(\gamma)/\pi}
\left(\frac{2a(x_{\cal P})}{\pi}\right)^3  \int dx_{12}dx_{02} W(x_{12},x_{02})
D(x_{12}) D(x_{02})
\label{sigtil}
\end{equation}
    where
\begin{equation}
D(x)= \int d^2r dz \Phi(r,z) r ln \frac{b^2}{rx} \exp[-\frac{a(x_{\cal P})}{2}
ln^2\frac{b^2}{rx}].  \label{dx}
\end{equation}
Using (\ref{W}) it is possible to perform the integrals in  (\ref{sigtil})
and obtain the following expression for the Mellin transform of $\sigma_d$
\begin{equation}
 \nonumber \tilde{\sigma}_d =  32 \pi \alpha^5 N_c <x^2(b,x_{\cal P})>^2 \\
\frac{1}{b^{2-\gamma}}  \frac {V(\gamma)}\gamma \
\beta^{-\alpha N_c \chi(\gamma)/\pi}
\label{sigtilde}
\end{equation}
where
\begin{eqnarray}
\nonumber <x^2(b,x_{\cal P})>= \int d^2r dz \Phi(r,z) \int \frac{dx}{x} x^2
\rho_1(r,b,x,x_{\cal P})= \\ = \frac{1}{4b} x_{\cal P}^{-\Delta_P}
(\frac{2a(x_{\cal P})}{\pi})^{\frac 32}   \int d^2r dz
\Phi(r,z) r \log(\frac br) \exp[-\frac{a(x_{\cal P})}{2} \log^2 (\frac br)].
\label{x2}
\end{eqnarray}
As seen from (\ref{x2}), the dimensionless quantity $<x^2>$ can be interpreted
as the density distribution for the average of the square
of the  transverse sizes of the dipoles inside the proton at a fixed impact
parameter b.
It
summarizes the whole information
about the proton wave function which is relevant for the process we consider.

To obtain $\sigma_d$ from (\ref{sigtilde}) one has to perform the inverse
Mellin transform
\begin{equation}
\sigma_d(\bar{r};b,;\beta,x_{\cal P})
 = \int_{c-i\infty}^{c+i\infty}\frac{d\gamma}{2\pi i}\
\bar{r}^{2 -\gamma}\  \tilde{\sigma}_d(
\gamma;b;\beta,x_{\cal P})
 \label{Hinv}
\end{equation}
with $c>0$. Inserting (\ref{sigtilde}) and (\ref{Hinv}) into (\ref{sigd}) and
using
\cite{nik}
(we neglect quark masses and the longitudinal cross-section)
\begin{equation}
\bar{\Phi} (r,z;Q^2) = \frac{N_c \alpha_{em}}{(2\pi)^2}  e_f^2
(z^2
+
(1-z)^2) \hat{Q}^2 K_1(\hat{Q}r) \label{fibar}
\end{equation}
with $\hat{Q}^2 = z(1-z)Q^2$ one can perform the integrations over z and r. The
result is
\begin{eqnarray}
\frac{\beta d\sigma}{d\beta d^2b}=
 \nonumber  \frac{32 \alpha_{em}}{ \pi}e^2_f \alpha^5 N_c^2 <x^2(b,x_{\cal
P})>^2
(bQ)^{-2} \\
 \int_{c-i\infty}^{c+i\infty}\frac{d\gamma}{2\pi i}\
(\frac{bQ}2)^{\gamma}
 (2-\gamma)^3 H(\gamma)\
\beta^{-\alpha N_c \chi(\gamma)/\pi} \label{sigfin}
\end{eqnarray}

Eq.(\ref{sigfin}) represents our final result for the diffractive
photon-nucleon cross-section at fixed impact parameter.

The diffractive structure function as defined, e.g., in \cite{h1,zeus} is
obtained from
(\ref{sigfin})
 using the relation
\begin{equation}
F_2^{D(4)}(Q^2,x_{\cal P},\beta,b) = \frac{Q^2}{4\pi^2\alpha_{em}}x_{\cal
P}^{-1}\frac{\beta d\sigma}{d\beta d^2b}.
\label{fd}
\end{equation}

Since b-dependence of the diffractive structure function is not
experimentally accessible, in the following we consider its integral over
$d^2b$
which is obviously equal to the integral over the whole range of the momentum
transfer $t$ to the target proton and thus measurable
\footnote{For structure functions, the relation between b-dependence and t-dependence is not
simple and goes beyond the scope of this work.}.
To integrate
 (\ref{sigfin}) over $d^2b$, however,  it is necessary to know the form of
$<x^2(b,x_{\cal P})>$ and thus the form of the proton wave function
$\Phi(r,z)$. Fortunately, in the limit $x_{\cal P} \rightarrow 0$, and $b \gg
r$,
$<x^2>$ is not very sensitive to this input and can be approximated as
\begin{equation}
<x^2(b,x_{cal P})>=
 \frac{r_0}{4b} x_{\cal P}^{-\Delta_P}
(\frac{2a(x_{\cal P})}{\pi})^{\frac 32}
 \log(\frac {b}{r_0}) \exp[-\frac{a(x_{\cal P})}{2} \log^2 (\frac {b}{r_0})].
\label{x20}
\end{equation}
where
\begin{equation}
r_0= \int d^2rdz \Phi(r,z) r.   \label{r0}
\end{equation}
Using (\ref{x20}) one can integrate (\ref{sigfin}) over $b$. Taking into
account (\ref{fd}) we obtain (\ref{Fdfin}).

3. To summarize, using the QCD dipole framework we have calculated the large
mass contribution to the process of diffraction dissociation of the virtual
photon at large $Q^2$ in the limit of very small $x_{\cal P}$. We find that the
diffractive structure function, when integrated over t, takes a particularly
simple factorizable form
\footnote{This is in contrast with the rather complicated behaviour at small t
found in \cite{t}.}.
However, the resulting "Pomeron flux factor" is modified by logarithmic
corrections
which lead to an effective intercept substantially lower than the one obtained
from the proton structure function. This should have clear experimental
consequences.

 Our calculation provides an explicit formula for the triple
(hard-)pomeron
coupling. In this context we find a rather large asymptotic (i.e. for
$\gamma = 1$) enhancement factor
as compared to the first order perturbative calculation. This means that the
Lipatov resummation is even more important here than in the total
cross-section \cite{catani}.

 Our results imply that the pattern of scaling violation at small $\beta$
should be similar to that observed for total cross section at small $x_{Bj}$.
It is important to realize, however, that this conclusion does not apply for
large values of $\beta$ (i.e. finite ratio $M^2/Q^2$). In this region of
$\beta$ the valence $q \bar{q}$ content of the pomeron should be taken into
account (see, e.g., \cite{gen} and references quoted there). This goes beyond
the scope of the present investigation.

{\bf Acknowledgments}

We would like to thank Al. Mueller, H. Navelet, Ch. Royon, G. Salam
and S. Wallon for many fruitful discussions. This work has been
supported by the exchange programme between the Polish and French
Academies of Sciences, the KBN grant (N0 2 P03B 083 08) and by PECO grant from
the EEC Programme "Human Capital and Mobility", Network "Physics at
High Energy Colliders" (Contract Nr: ERBICIPDCT940613).


\begin{thebibliography}{99}
\bibitem{h1}
H1 coll., T.Ahmed et al. \ {\it Phys.Letters} {\bf B348} (1995) 681.
\bibitem{zeus}
ZEUS coll., M.Derrick et al. DESY Preprint 95-093.
\bibitem{Ingelman}
G.Ingelman and P.Schlein, {\it Phys. Lett.} {\bf B152} (1985) 256.
\bibitem{theory}
W.Buchmuller and A. Hebecker, {\it Phys. Lett.} {\bf B355} (1995) 573;
A.Edin, J.Rathsman and G. Ingelman, DESY 95-163,TSL/ISV-95-0125.
\bibitem{muel}
A.H.Mueller, {\it Nucl. Phys.} {\bf B415} (1994) 373.
\bibitem{nik}
N.N.Nikolaev and B.G.Zakharov, {\it Zeit. fur. Phys.} {\bf C49} (1991) 607;
{\it ibid.} {\bf C64} (1994) 651.
\bibitem{muela}
A.H.Mueller and B.Patel, {\it Nucl. Phys.} {\bf B425} (1994) 471.
\bibitem{lip}
V.S.Fadin, E.A.Kuraev and L.N.Lipatov   {\it Phys. lett.} {\bf B60} (1975)
50; I.I.Balitsky and L.N.Lipatov, {\it Sov.J.Nucl.Phys.} {\bf 15} (1978) 438.
\bibitem{zeusa}
ZEUS collab. M.Derrick et al., {\it Phys. Lett.} {\bf B316} (1993) 201.
\bibitem{h1a}
H1 collab. I.Abt et al., {\it Nucl. Phys. } {\bf B407} (1993) 515,
{\it Phys. Lett.} {\bf B321} (1994) 161.
\bibitem{NPR}
H.Navelet, R.Peschanski and Ch. Royon, hep-ph/9508259, to appear soon in {\it Phys. Lett.}
\bibitem{catani}
S.Catani, M. Ciafaloni and F. Hautmann, {\it Phys. Lett.} {\bf B242} (1990) 97,
{\it Nucl. Phys.} {\bf B366} (1991) 135.
\bibitem{good}
M.L.Good and W.D.Walker, {\it Phys. Rev.} {\bf 120} (1960) 1857.
\bibitem{miet}
M.T.Miettinen and J.Pumplin, {\it Phys.Rev.} {\bf D18} (1978) 1696.
\bibitem{muelb}
A.H.Mueller, {Nucl. Phys.} {\bf B} 437 (1995) 107.
\bibitem{BP1}
A.Bialas and R.Peschanski, {\it Phys.Lett.} {\bf B355} (1995) 301.
\bibitem{t}
M.Braun, Preprint SPbU IP 1995/10, hep-ph/9506245; 
J.Bartels, H.Lotter and M.Wusthoff, DESY preprint, hep-ph/9501314.
\bibitem{gen}
M.Genovese, N.N.Nikolaev and B.G.Zakharov, Preprint KFA-IKP(Th)-1994-37 (unpublished);
K.Golec-Biernat and J.Kwiecinski, {\it Phys. Lett.} {\bf B353} (1995) 329.
\end{thebibliography}
\end{document}